\documentclass[letter,twocolumn]{jpsj2} 
\usepackage{txfonts}

\def\runauthor{Hiroaki \textsc{Kusunose} and Yoshio \textsc{Kuramoto}}

\title{
Evidence for Octupole Order in Ce$_{0.7}$La$_{0.3}$B$_6$ from Resonant X-ray Scattering 
}

\author{
Hiroaki \textsc{Kusunose}\thanks{E-mail: kusu@cmpt.phys.tohoku.ac.jp}
and Yoshio \textsc{Kuramoto}
}

\inst{
Department of Physics, Tohoku University, Sendai 980-8578
}

\abst{
The azimuthal angle dependence observed in the resonant X-ray scattering in phase IV of  Ce$_{0.7}$La$_{0.3}$B$_6$ is analyzed theoretically.
It is shown that the peculiar angle dependence observed in the $E2$ channel is consistent with 
the $\Gamma_{5u}$-type octupole order
with principal axis along (111) and equivalent directions. 
Under the assumption that the four equivalent octupole domains are nearly equally populated in the sample, 
the observed angle dependences are reproduced by calculation for both $\sigma\sigma'$ 
and $\sigma\pi'$ polarizations.
The calculation for various symmetries of order parameters
excludes unambiguously other order parameters than the $\Gamma_{5u}$-type octupole. 
}

\kword{
octupole moment, resonant X-ray scattering, rare-earth hexaboride
}

\begin{document}
\maketitle

Octupole orders have attracted growing interest recently in relation to unusual properties found in NpO$_2$ \cite{Santini00,Paixao02,Caciuffo03},  
URu$_2$Si$_2$ \cite{Kiss03}, and 
Ce$_x$La$_{1-x}$B$_6$ \cite{Kuramoto00,Kusunose01,Kubo04}.
In particular, 
Ce$_x$La$_{1-x}$B$_6$ exhibits a rich variety of the $H$-$T$ phase diagram, called I--IV\cite{Hiroi97,Tayama97}.
The importance of the octupole degrees of freedom was first recognized by Sakai and co-workers\cite{Sakai97,Shiina98} in resolving the inconsistency between NMR and the neutron scattering measurements in the antiferro-quadrupole phase II.
Soon after their work,  possible relevance of the $\Gamma_{5u}$-type octupole to the magnetic phase III and III' was pointed out\cite{Kusunose01}.
For consistent understanding of the overall phase diagram, this type of octupole was also suggested as a hidden order parameter of phase IV\cite{Kusunose01}.
The prediction of the accompanied lattice distortion along the [111] axis is indeed observed\cite{Akatsu03}.
The model of the $\Gamma_{5u}$-type octupole order 
explains most of the characteristic behaviors in phase IV\cite{Kubo04}, such as the cusp-like behavior in the uniform susceptibility\cite{Tayama97}, no significant magnetic reflection in neutron scattering\cite{Iwasa03}, and the huge softening of $C_{44}$ mode in the strain susceptibility\cite{Suzuki98} as well as the spontaneous lattice distortion along [111]\cite{Akatsu03}.
However, there has been no direct evidence from diffraction experiment.

Resonant X-ray scattering (RXS) measurement has been performed recently to obtain a direct evidence of the order parameter symmetry in phase IV \cite{Mannix05}.
The RXS has already proven its efficiency in various systems such as LaMnO$_3$\cite{Murakami98}, DyB$_2$C$_2$\cite{Hirota00,Matsumura02,Tanaka04} and NpO$_2$\cite{Paixao02,Caciuffo03}.
The measurement was done for Ce$_{0.7}$La$_{0.3}$B$_6$ below $T_{\rm IV}=1.5$ K, using the $L_2$ absorption edge in the Ce atom\cite{Mannix05}.
They measured the azimuthal angle dependences of the 
superlattice Bragg reflections at 
${\mib q}=(3/2,3/2,3/2)$ 
with the rotation axis normal to the [111] surface.
Both the non-rotated ($\sigma$-$\sigma'$), and the rotated ($\sigma$-$\pi'$) polarization channels have been measured at the electric quadrupole (E2) resonance.
In the $\sigma$-$\sigma'$ channel, the scattering intensity exhibits six-fold oscillation, which indicates the occurrence of the electronic order with three-fold symmetry along [111].
In contrast, the $\sigma$-$\pi'$ channel shows three-fold oscillation, which remains to be clarified.
It is not clear to us how the fitting in ref.\citen{Mannix05} is justfied microscopically.

In this Letter, we demonstrate that the $\Gamma_{5u}$-type octupole order
can reproduce the azimuthal angle dependences both in the $\sigma$-$\sigma'$ and $\sigma$-$\pi'$ channels.
The three-fold oscillation is explained in terms of domains whose principal axis 
is other than [111].  These domains give the scattering amplitude which has both
three-fold oscillatory component and the axially symmetric one.  As a result, their cross product contributing the intensity gives the three-fold symmetry.
We adopt a concise formalism for the RXS amplitude proposed by Lovesey and Balcar\cite{Lovesey96a,Lovesey96b}, which assumes the spherical symmetry in the intermediate states.
Our results of calculation are compiled into tables, 
which allow us to analyze various multipole orders quite easily.
With an appropriate rotation to obtain domains with different principal axes,
we summarize the azimuthal angle dependences for all possible order parameters.
The result excludes unambiguously the other symmetries than the $\Gamma_{5u}$ octupole.

The RXS amplitude per unit cell
is given by the standard formula: \cite{Lovesey96a,Lovesey96b} 
\begin{equation}
F_{\rm reso}=-\frac{\Delta^2}{\hbar^2c^2}\sum_{m}\frac{W_{fi}^{(m)}}{\hbar\omega-\Delta+i\Gamma/2},
\end{equation}
where we have neglected the energy distribution of the intermediate states $(m)$,  and of their width $\Gamma$.
In the RXS, the energy of the incident photon $\hbar\omega$ is tuned close to the absorption edge $\Delta$ of the relevant atom.
The intensity is proportional to 
\begin{multline}
W_{fi}^{(m)}=\langle f| {\mib \epsilon}'\cdot{\mib P}|m\rangle\langle m|{\mib \epsilon}\cdot{\mib P}|i\rangle
\\
+\langle f| {\rm Tr}(\hat{X}'\cdot\hat{Q})|m\rangle\langle m|{\rm Tr}(\hat{X}\cdot\hat{Q}|i\rangle,
\end{multline}
where the first term in the right-hand side comes 
from the electric dipole (E1) transition, and the second term from the quadrupole (E2) transition.
Here we have defined the dipole and the quadrupole operators in an atom, $P^\alpha=e\sum_{n}^{\rm atom}r_n^\alpha$ and $\hat{Q}_{\alpha\beta}=e\sum_n^{\rm atom}r_n^\alpha r_n^\beta/2$, and the matrix made from the photon wave and polarization vectors, $\hat{X}_{\alpha\beta}=k^\alpha \epsilon^\beta/2$.
The prime represents quantities concerning the scattered radiation.

Following Lovesey and Balcar\cite{Lovesey96a,Lovesey96b}, we assume that the intermediate states can be described by an angular momentum of a core hole.
Then, the summation in eq.~(1) can be done with the help of the Wigner-Eckart theorem, and we are left with two scalar products of the spherical tensors:
\begin{subequations}
\begin{align}
&\sum_m W^{(m)}_{fi}=Z_{fi}^{E1}+Z_{fi}^{E2},
\nonumber\\
&Z_{fi}^{E1}=\sum_{p=0}^2 c^{\rm E1}_p\sum_{q=-p}^pT^{(p)}_q(M,M')K^{(p)*}_q({\mib\epsilon}',{\mib\epsilon}),
\\
&
Z_{fi}^{E2}=\sum_{p=0}^4 c^{\rm E2}_p\sum_{q=-p}^pT^{(p)}_q(M,M')H^{(p)*}_q(\hat{Q}',\hat{Q}),
\end{align}
\end{subequations}
where $T^{(p)}_q$ represents the multipole degrees of freedom in the $f$ shell.
It is related to the spherical harmonics as $T^{(p)}_q=\sqrt{4\pi/(2p+1)}r^p Y_{pq}$ where $r$ is the radial coordinate.
The operator equivalents of $T^{(p)}_q(M,M')$ can be obtained by replacing the coordinate ${\mib r}$ with its symmetric product of the total angular momentum ${\mib J}$, e.g., $xy^2\to (J_xJ_y^2+J_yJ_xJ_y+J_y^2J_x)/3$.
The information of ${\mib k}$ and ${\mib\epsilon}$ is summarized concisely in the corresponding spherical tensors,
\begin{subequations}
\begin{align}
&K^{(p)}_q=i^p\sum_{\alpha\beta=-1}^1\epsilon_\alpha'\epsilon_\beta \langle 1\alpha,1\beta|pq\rangle,
\\
&H^{(p)}_q=i^p\sum_{\alpha\beta=-2}^2h'_\alpha h_\beta\langle 2\alpha,2\beta|pq\rangle,
\\
&h_\alpha=\sum_{\gamma\delta=-1}^1\epsilon_\gamma k_\delta \langle 1\gamma,1\delta|2\alpha\rangle,
\end{align}
\end{subequations}
where $\langle j_1m_1,j_2m_2|jm\rangle$ is the Clebsch-Grodan coefficient.
An ordinary vector ${\mib v}$ in the cartesian coordinate is related to the 1st rank spherical tensor as $v^{(1)}_{\pm 1}=\mp(v_x+iv_y)/\sqrt{2}$ and $v^{(1)}_0=v_z$.
For all spherical tensors in this paper, we use the phase convention 
$[S^{(p)}_q]^*=(-1)^qS^{(p)}_{-q}$.\cite{phase-conv}

The operator ${\mib J}$ is the axial vector and has an odd parity under time reversal\cite{Chatterjee83}.
As a consequence, within the same manifold of the angular momentum, an odd-rank tensor operator describes a magnetic multipole, which can be coupled to an axial tensor with time-reversal odd, such as $i{\mib\epsilon}'\times{\mib\epsilon}$, while an even-rank tensor operator represents an electric multipole which appears only with the polar tensor with time-reversal even.
When a spontaneous order of $T^{(p)}_{q}$ multipole occurs, i.e., $\langle T^{(p)}_q\rangle\ne0$, the azimuthal angle dependence of the corresponding $H^{(p)}_q$ will appear in the RXS amplitude.
The scalar form of eq.~(1) guarantees that arbitrary linear combination of $T^{(p)}_q$ couples with the complex conjugate of the same linear combination of $H^{(p)}_q$.
The scalar property 
is a great advantage in considering the case where 
the ordered moment is directed to a direction different from the $z$ axis.
This situation will be discussed in more detail later.
Since the explicit form of the coefficient $c_p^{En}\  (n=1,2)$ depends on the representation of multiplet, i.e., the Russell-Saunders or the $j$-$j$ coupling, 
we regard it as a fitting parameter.
For further details, we refer to the papers by Lovesey and Balcar\cite{Lovesey96a,Lovesey96b}.

\begin{figure}
\begin{center}
\includegraphics[width=7cm]{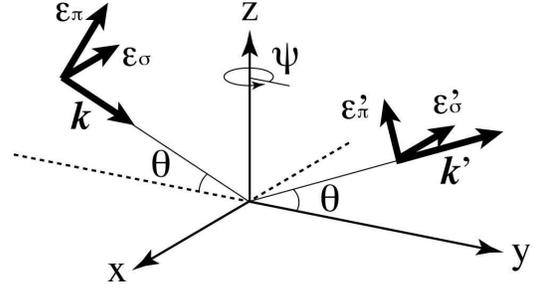}
\caption{ The coordinate system corresponding to the RXS experiment.
The azimuthal angle $\psi$ dependence is obtained by rotating the 
 ${\mib r}=(x,y,z)$ coordinate system around the $z$ axis relative to the photon ${\mib k}$ vector, i.e., ${\mib k}$ is in the $yz$ plane at $\psi=0$. 
 The $\sigma$ and $\pi$ polarization vectors are also shown (${\mib\epsilon}_{\sigma}\times{\mib\epsilon}_{\pi}={\mib k}/|{\mib k}|$).}
\end{center}
\end{figure}

\begin{table}[t]
\caption{The angle dependences of the RXS {\it amplitude} in the non-rotated ($\sigma$-$\sigma'$) and the rotated ($\sigma$-$\pi'$) polarization channels for the E1 transition. 
The common coordinate systems for the scattering and the crystal is used. The definition of the cartesian spherical tensor $K^{(p)}_{cq,sq}$ is given in the text.}
\begin{tabular}{ccc}
\\ \hline\hline
 & $\sigma\sigma'$ & $\sigma\pi'$ \\ \hline
$K^{(0)}_0$  &  $-\frac{1}{\sqrt{3}}$ & 0 \\ \hline 
$K^{(1)}_{c1}$ & 0 & $\frac{1}{\sqrt{2}}\cos\theta\sin\psi$ \\
$K^{(1)}_{s1}$ & 0 & $\frac{1}{\sqrt{2}}\cos\theta\cos\psi$ \\
$K^{(1)}_0$ & 0 & $\frac{1}{\sqrt{2}}\sin\theta$ \\ \hline 
$K^{(2)}_{c2}$ & $-\frac{1}{\sqrt{2}}\cos(2\psi)$ & $-\frac{1}{\sqrt{2}}\sin\theta\sin(2\psi)$ \\
$K^{(2)}_{c1}$ & 0 & $\frac{1}{\sqrt{2}}\cos\theta\cos\psi$ \\
$K^{(2)}_{s2}$ & $\frac{1}{\sqrt{2}}\sin(2\psi)$ & $-\frac{1}{\sqrt{2}}\sin\theta\cos(2\psi)$ \\
$K^{(2)}_{s1}$ & 0 & $-\frac{1}{\sqrt{2}}\cos\theta\sin\psi$ \\
$K^{(2)}_0$ & $\frac{1}{\sqrt{6}}$ & 0 \\ \hline\hline
\end{tabular}
\end{table}

\begin{table}[t]
\caption{The angle dependences of the RXS {\it amplitude} for the E2 transition.}
\begin{tabular}{ccc}
\\ \hline\hline
 & $\sigma\sigma'$ & $\sigma\pi'$ \\ \hline
$H^{(0)}_0$ & $\frac{1}{2\sqrt{5}}\cos(2\theta)$ & 0 \\ \hline
$H^{(1)}_{c1}$ & $\frac{1}{2\sqrt{10}}\sin(2\theta)\cos\psi$ & $\frac{1}{2\sqrt{10}}\cos(3\theta)\sin\psi$ \\
$H^{(1)}_{s1}$ & $-\frac{1}{2\sqrt{10}}\sin(2\theta)\sin\psi$ & $\frac{1}{2\sqrt{10}}\cos(3\theta)\cos\psi$ \\
$H^{(1)}_0$ & 0 & $\frac{1}{2\sqrt{10}}\sin(3\theta)$ \\ \hline
$H^{(2)}_{c2}$ & $-\frac{1}{2}\sqrt{\frac{3}{14}}\sin^2\theta\cos(2\psi)$ & $\frac{1}{2}\sqrt{\frac{3}{14}}\sin(3\theta)\sin(2\psi)$ \\
$H^{(2)}_{c1}$ & 0 & $-\frac{1}{2}\sqrt{\frac{3}{14}}\cos(3\theta)\cos\psi$ \\
$H^{(2)}_{s2}$ & $\frac{1}{2}\sqrt{\frac{3}{14}}\sin^2\theta\sin(2\psi)$ & $\frac{1}{2}\sqrt{\frac{3}{14}}\sin(3\theta)\cos(2\psi)$ \\
$H^{(2)}_{s1}$ & 0 & $\frac{1}{2}\sqrt{\frac{3}{14}}\cos(3\theta)\sin\psi$ \\
$H^{(2)}_0$ & $-\frac{1}{4\sqrt{14}}(3+\cos(2\theta))$ & 0 \\ \hline
$H^{(3)}_{c3}$ & $\frac{1}{4}\sin(2\theta)\cos(3\psi)$ & $-\frac{1}{16}(\cos\theta+3\cos(3\theta))\sin(3\psi)$ \\
$H^{(3)}_{c2}$ & 0 & $-\frac{1}{2}\sqrt{\frac{3}{2}}\cos^2\theta\sin\theta\cos(2\psi)$ \\
$H^{(3)}_{c1}$ & $-\frac{1}{4}\sqrt{\frac{3}{5}}\sin(2\theta)\cos\psi$ & $\frac{1}{8}\sqrt{\frac{3}{5}}\cos\theta(\cos(2\theta)-3)\sin\psi$ \\
$H^{(3)}_{s3}$ & $-\frac{1}{4}\sin(2\theta)\sin(3\psi)$ & $-\frac{1}{16}(\cos\theta+3\cos(3\theta))\cos(3\psi)$ \\
$H^{(3)}_{s2}$ & 0 & $\frac{1}{2}\sqrt{\frac{3}{2}}\cos^2\theta\sin\theta\sin(2\psi)$ \\
$H^{(3)}_{s1}$ & $\frac{1}{4}\sqrt{\frac{3}{5}}\sin(2\theta)\sin\psi$ & $\frac{1}{8}\sqrt{\frac{3}{5}}\cos\theta(\cos(2\theta)-3)\cos\psi$ \\
$H^{(3)}_0$ & 0  & $\frac{1}{4\sqrt{10}}\sin\theta(3\cos(2\theta)-1)$ \\ \hline
$H^{(4)}_{c4}$ & $-\frac{1}{2\sqrt{2}}\cos^2\theta\cos(4\psi)$ & $-\frac{1}{2\sqrt{2}}\sin\theta\cos^2\theta\sin(4\psi)$ \\
$H^{(4)}_{c3}$ & 0 & $\frac{1}{4}\cos^3\theta\cos(3\psi)$ \\
$H^{(4)}_{c2}$ & $-\frac{1}{\sqrt{14}}\sin^2\theta\cos(2\psi)$ & $\frac{1}{4\sqrt{14}}\sin\theta(\cos(2\theta)-3)\sin(2\psi)$ \\
$H^{(4)}_{c1}$ & 0 & $-\frac{1}{8\sqrt{7}}\cos\theta(3\cos(2\theta)-5)\cos\psi$ \\
$H^{(4)}_{s4}$ & $\frac{1}{2\sqrt{2}}\cos^2\theta\sin(4\psi)$ & $-\frac{1}{2\sqrt{2}}\sin\theta\cos^2\theta\cos(4\psi)$ \\
$H^{(4)}_{s3}$ & 0 & $-\frac{1}{4}\cos^3\theta\sin(3\psi)$ \\
$H^{(4)}_{s2}$ & $\frac{1}{\sqrt{14}}\sin^2\theta\sin(2\psi)$ & $\frac{1}{4\sqrt{14}}\sin\theta(\cos(2\theta)-3)\cos(2\psi)$ \\
$H^{(4)}_{s1}$ & 0 & $\frac{1}{8\sqrt{7}}\cos\theta(3\cos(2\theta)-5)\sin\psi$ \\
$H^{(4)}_0$ & $\frac{1}{4\sqrt{70}}(5-3\cos(2\theta))$ & 0 \\ \hline\hline
\end{tabular}
\end{table}

The experimental geometry of the scattering is shown in Fig.~1 with the
coordinate system ${\mib r}=(x,y,z)$.
The azimuthal angle $\psi$ dependence is obtained by rotating the ${\mib r}$ coordinate system around the $z$ axis relative to the photon ${\mib k}$ vector.
The ${\mib k}$ vector is then given by ${\mib k}=k(\cos\theta\sin\psi,\cos\theta\cos\psi,\sin\theta)$.
Tables ~I and II summarize the Bragg and the azimuthal angle dependences of $K^{(p)}_q$ and $H^{(p)}_q$ 
for the case where the crystal coordinate ${\mib R}=(X,Y,Z)$
is the same as the scattering coordinate ${\mib r}$.
Here we have used the real cartesian tensors for $q>0$ defined as
\begin{subequations}
\begin{align}
&
K^{(p)}_{cq}=\frac{(-1)^q}{\sqrt{2}}\left[ K^{(p)}_q + K^{(p)*}_q \right],
\\
&
K^{(p)}_{sq}=\frac{(-1)^q}{\sqrt{2}i}\left[ K^{(p)}_q - K^{(p)*}_q \right].
\end{align}
\end{subequations}
A similar definition is applied to $H^{(p)}_q$ as well.
Due to the choice of ${\mib\epsilon}_\sigma\parallel x$ for $\psi=0$, the $\psi$ dependence in the $\sigma$-$\sigma'$ channel is the same as that of the spherical harmonics.
On the other hand, the change of the photon helicity in the $\sigma$-$\pi'$ channel gives an additional phase factor $\pi/2$ in the $\psi$ dependence.
For an occurrence of spontaneous multipole order, we obtain immediately the $\psi$ dependence of the RXS amplitude from Tables~I and II for the E1 and the E2 transitions, respectively,
as long as the principal axis of the order parameter is parallel to the $z$ axis.

Now, let us discuss the azimuthal angle dependence of phase IV in Ce$_{0.7}$La$_{0.3}$B$_6$.
The E1 scattering shows a short-range magnetic order, which may be induced from the octupole order together with some crystalline disorder.
In the following, we restrict our discussion to the E2 transition, since it provides the direct information of the octupoles.
In order to take the [111] direction as the rotation axis $z$, we set
\begin{equation}
\left(\begin{array}{c}X\\ Y\\ Z\end{array}\right)
=
\left(
\begin{array}{ccc}
1/\sqrt{2} & 1/\sqrt{6} & 1/\sqrt{3} \\
-1/\sqrt{2} & 1/\sqrt{6} & 1/\sqrt{3} \\
0 & -2/\sqrt{6} & 1/\sqrt{3}
\end{array}
\right)
\left(\begin{array}{c}x\\ y\\ z\end{array}\right).
\end{equation}
Then, $[11\bar{2}]$ direction is parallel to 
$y$,  and ${\mib k}$ at $\psi=0$ is in the $yz$ plane.
The origin of $\psi$ is the same as the one used by Mannix {\it et al.}\cite{Mannix05}.

The most promising candidate of the order parameter for phase IV is the $\Gamma_{5u}$-type octupole with the [111] principal axis:
\begin{multline}
T^{[111]}_{5u}({\mib R})=\frac{1}{\sqrt{3}}\biggl[\frac{\sqrt{15}}{2}
\biggl\{X(Y^2-Y^2)+Y(Z^2-X^2)
\\
+Z(X^2-Y^2)\biggr\}\biggr]=\frac{\sqrt{5}}{2}(X-Y)(Y-Z)(Z-X).
\end{multline}
In the scattering coordinate, it has the form
\begin{equation}
T^{[111]}_{5u}({\mib r})=\frac{1}{2}\sqrt{\frac{5}{2}}x(x^2-3y^2)=T^{(3)}_{c3}({\mib r}).
\end{equation}
Thus, from the row of $H^{(3)}_{c3}$ in Table~II, we obtain immediately the azimuthal angle dependences of the $T^{[111]}_{5u}$ order as
\begin{subequations}
\begin{align}
&
|f^{[111]}_{5u}(\sigma\sigma')|^2=\frac{1}{16}\sin^2(2\theta)\cos^2(3\psi),
\\
&
|f^{[111]}_{5u}(\sigma\pi')|^2=\frac{1}{256}[\cos\theta+3\cos(3\theta)]^2\sin^2(3\psi).
\end{align}
\end{subequations}

\begin{table}[t]
\caption{The angle dependence of the RXS {\it intensity} for possible multipoles in three-fold axis. The rotation axis is [111], and ${\mib k}$ is in the $[111]$-$[11\bar{2}]$ plane at $\psi=0$.}
\begin{tabular}{ccc}
\\ \hline\hline
$p$-$\Gamma$ &  E2 $\sigma\sigma'$ & E2 $\sigma\pi'$ \\ \hline
1-4u & 0 & $\frac{1}{40}\sin^2(3\theta)$ \\ 
2-5g & $\frac{1}{224}(3+\cos(2\theta))^2$ & 0 \\
3-2u & $\frac{1}{36}\sin^2(2\theta)\sin^2(3\psi)$ & $\frac{1}{144}(3\cos(2\theta)-1)^2\cos^2\theta\times$ \\
 & & $\mbox{\hspace{0.3cm}}\times\left[\frac{1}{\sqrt{2}}\tan\theta-\cos(3\psi)\right]^2$ \\
3-4u & $\frac{1}{36}\sin^2(2\theta)\sin^2(3\psi)$ & $\frac{1}{144}(3\cos(2\theta)-1)^2\cos^2\theta\times$ \\
& & $\mbox{\hspace{0.3cm}}\times\left[\frac{1}{\sqrt{2}}\tan\theta+\cos(3\psi)\right]^2$ \\
3-5u & $\frac{1}{16}\sin^2(2\theta)\cos^2(3\psi)$ & $\frac{1}{256}(\cos\theta+3\cos(3\theta))^2\sin^2(3\psi)$ \\
4-4g & 0 & $\frac{1}{16}\cos^6\theta\cos^2(3\psi)$ \\
4-5g & $\frac{1}{1512}(5-3\cos(2\theta))^2$ & $\frac{1}{1296}\cos^6\theta\sin^2(3\psi)$ \\
\hline\hline
\end{tabular}
\end{table}

Similarly, we derive the angle dependences of other
multipoles with [111] as the principal axis.
The expressions of the multipoles with components $p$-$\Gamma$ 
are given by
\begin{subequations}
\begin{align}
&T^{[111]}_{1-4u}=T^{(1)}_{0}({\mib r}), \\
&T^{[111]}_{2-5g}=T^{(2)}_{0}({\mib r}), \\
&T^{[111]}_{3-2u}=\frac{2}{3}T^{(3)}_{s3}({\mib r})+\frac{\sqrt{5}}{3}T^{(3)}_{0}({\mib r}), \\
&T^{[111]}_{3-4u}=\frac{\sqrt{5}}{3}T^{(3)}_{s3}({\mib r})-\frac{2}{3}T^{(3)}_{0}({\mib r}), \\
&T^{[111]}_{4-4g}=-T^{(4)}_{c3}({\mib r}), \\
&T^{[111]}_{4-5g}=\frac{2\sqrt{15}}{9}T^{(4)}_{0}({\mib r})+\frac{\sqrt{21}}{9}T^{(4)}_{s3}({\mib r}), \end{align}
\end{subequations}
and their angle dependences are summarized in Table~III.

The averaged scattering intensity from the other domains is calculated as
\begin{subequations}
\begin{align}
&
|f^{\ne[111]}_{5u}(\sigma\sigma')|^2=\frac{1}{144}\sin^2(2\theta)[\cos^2(3\psi)+2],
\\
&
|f^{\ne[111]}_{5u}(\sigma\pi')|^2=\frac{1}{4608}[
5(\cos(3\theta)-5\cos\theta)^2
\nonumber\\&\mbox{\hspace{1cm}}
-(3\cos(3\theta)+\cos\theta)\cos(6\psi)
\nonumber\\&\mbox{\hspace{1cm}}
+32\sqrt{2}(\sin(3\theta)-7\sin\theta)\cos^3\theta\cos(3\psi).
\end{align}
\end{subequations}
Among all multipoles accessible by the RXS, only the $\Gamma_{5u}$-type octupole yields the maximum at $\psi=0$ in the $\sigma$-$\sigma'$ channel, which is consistent with the observed oscillation\cite{Mannix05} as shown in Fig.~2.
In the $\sigma$-$\pi'$ channel, on the other hand, 
the contribution from the other three equivalent domains is much larger than that from [111] domain, and the threefold component appears with the maximum at $\psi=0$.
To the contrary, the [111] domain predominates over the other domains in the $\sigma$-$\sigma'$ channel.

Finally, we show the result of the fitting for the observed data 
taken from Mannix {\it et al.}\cite{Mannix05} in Fig.~2.
We have used the fitting formula for the
scattering channel $\alpha \ (=\sigma, \pi)$: 
\begin{equation}
I_\alpha=A_\alpha\left[
w|f_{5u}^{[111]}(\alpha)|^2+(1-w)|f_{5u}^{\ne[111]}(\alpha)|^2\right],
\label{fit}
\end{equation}
with the scattering angle $\theta=-39^\circ$ from the incident photon energy and the scattering wave vector ${\mib q}$\cite{Mannix05}.
We assume that the four equivalent domains are equally populated, i.e., $w=1/4$.
The calculated intensity is insensitive to slight change of $w$.
The intensity factors are chosen as
$A_{\sigma\sigma'}=120$, $A_{\sigma\pi'}=70$. 
If the ideal experimental setup were achieved, the intensity factors in the two scattering channels should be the same.
The discrepancy of the two intensity factors in the fitting may be ascribed to slightly different setup for different polarizations, and/or extrinsic background from non-resonant contribution.
The fitting with these intensity factors reproduces semi-quantitatively the observed data.
Thus, we conclude that the $\Gamma_{5u}$-type octupole in three-fold axis with the four equivalent domains is the order parameter of phase IV in Ce$_{0.7}$La$_{0.3}$B$_6$.

\begin{figure}
\begin{center}
\includegraphics[width=8cm]{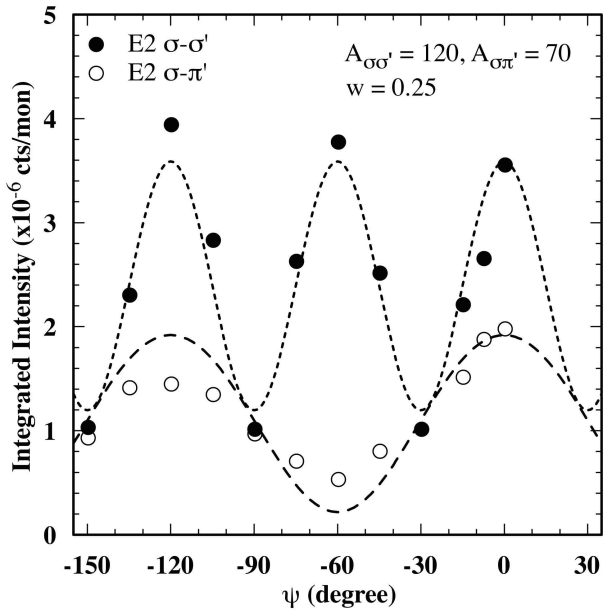}
\caption{The azimuthal angle dependence. The symbols are taken from Mannix {\it et al.}\cite{Mannix05}. The dotted and the dashed lines show the intensities from the $\Gamma_{5u}$-type octupole order described by eq.~(\ref{fit}) in the text.}
\end{center}
\end{figure}

In summary, on the basis of the formalism by Lovesey and Balcar, we have derived
azimuthal angle dependences for all multipole orders.
The appropriate rotation, which transforms original expressions of irreducible tensors into the scattering coordinate system, allows us to extract the azimuthal angle dependences for all possible candidates of the order parameters for phase IV in Ce$_{0.7}$La$_{0.3}$B$_6$.
In the non-rotated $\sigma$-$\sigma'$ scattering channel, only the $\Gamma_{5u}$-type octupole in three-fold principal axis yields consistent six-fold oscillatory behavior with the maximum at $\psi=0$.
On the other hand, in the rotated $\sigma$-$\pi'$ channel, the contribution from the other three equivalent domains predominates over the contribution from the [111] domain, showing three-fold oscillation with the maximum at $\psi=0$.
Therefore, we conclude that the $\Gamma_{5u}$-type octupole in three-fold axis with the four equivalent domains is the order parameter of phase IV in Ce$_{0.7}$La$_{0.3}$B$_6$.

We would like to thank D. Mannix, N. Bernhoeft, and Y. Tanaka for showing us the experimental data prior to publication.
We are also indebted to S. Lovesey, K. Katsumata, Y. Murakami, T. Matsumura and S. Ishihara for fruitful discussions.
This work was supported partly by 
Grants-in-Aid for Scientific Research on Priority Area ``Skutterudite" , and 
for Scientific Research (B)15340105 
of the Ministry of Education, Culture, Sports, Science and Technology, Japan.


\begin{thebibliography}{99}
\bibitem{phase-conv} In the papers by Lovesey and Balcar, they use alternative phase convention such that $[S^{(p)}_q]^*=(-1)^{p-q}S^{(p)}_{-q}$. With our definition, the $q=0$ component of the odd-rank tensors becomes real rather than pure imaginary.
\end{thebibliography}
\end{document}